# Electron-phonon coupling in cuprate and iron-based superconductors revealed by Raman scattering

An-min Zhang and Qing-ming Zhang[*]

Department of Physics, Renmin University, Beijing 100872, P. R. China

**Abstract**

Electron-phonon coupling (EPC) is one of the most common and fundamental interactions in solids. It not only dominates many basic dynamic processes like resistivity, thermal conductivity etc, but also provides the pairing glue in conventional superconductors. But in high-temperature superconductors (HTSC), it is still controversial whether or not EPC is in favor of paring. Despite the controversies, many experiments have provided clear evidence for EPC in HTSC. In this paper, we briefly review EPC in cuprate and iron-based superconducting systems revealed by Raman scattering. We introduce how to extract the coupling information through phonon lineshape. Then we discuss the strength of EPC in different HTSC systems and possible factors affecting the strength. The comparative study between Raman phonon theories and experiments allows us to gain insight into some crucial electronic properties, especially superconductivity. Finally we summarize and compare EPC in the two existing HTSC systems, and discuss what role it may play in HTSC.

---

[*] qmzhang@ruc.edu.cn

# Outline

## I. Introduction

Superconductivity was discovered in mercury in 1911 by Kamerling Onnes [1]. It was successively observed in many metals and alloys. But the superconducting (SC) mechanism was not well understood for quite a long time after the first discovery. Gortor and Casimir proposed a phenomenological two-fluid model in 1934 to explain zero resistivity [2], in which electrons are divided into two categories: normal electrons and SC ones carrying current without resistance. Afterwards, London equations were proposed to describe electromagnetic properties of a superconductor [3], which provides a beautiful explanation for Meissner diamagnetic behavior. In 1950, the phenomenological Ginzburg-Landau equation was raised based on a general Landau's second-order phase transition theory, which was considered to be a huge success in vortex dynamics and quantitative description of SC transition. Bardeen, Cooper and Schrieffer (BCS) proposed a microscopic pairing mechanism based on effective electron-phonon coupling [4], which eventually established a solid foundation for conventional superconductivity.

When Müller and Bednorz reported superconductivity at 35 K in layered compound $La_{2-x}Ba_xCuO_4$ in 1986 [5], the new era of HTSC was launched. The temperature barrier of 77 K liquid-nitrogen boiling point was soon broken [6, 7]. The record of the highest SC temperature was realized in HgBaCaCuO system under high pressure [8]. In the early stage of HTSC study, many researchers have attempted to understand the HTSC mechanism in the framework of BCS theories. But the small isotope effect revealed by experiments [9-12] failed to support the conventional EPC scenario. On the other hand, it was confirmed that cooper pairs are still the carriers of SC current in HTSC [13-15]. It means that a new pairing glue is needed for the HTSC systems, such as magnon [16], exciton [17, 18], polaron [19], bi-polaron [20, 21], antiferromagnetic (AF) fluctuations [22, 23], etc. At the same time, a strong EPC in cuprate HTSC was reported by a lot of experiments, which implies that EPC plays an important role in pairing, though not a dominant one. EPC is observed not only in Raman channel, but also in other measurements like neutron scattering and angle-resolved photoemission spectroscopy (ARPES), which has been comprehensively reviewed in Refs. 24-29. After over two decades, the discovery of high-Tc superconductivity in iron-based systems in 2008 was a great breakthrough in the HTSC study. It is generally expected that many unresolved problems in cuprates, including EPC, may be answered comparison with iron pnictides. As one of the most fundamental techniques in condensed matter physics, Raman scattering shows many unique advantages in EPC study. Through the changes of Raman phonons in lineshapes, widths and frequencies due to EPC, we can conveniently study the renormalization of electronic self-energy and precisely estimate the EPC strength, and hence explore its relation to superconductivity.

The reviews on electronic, phononic and magnon properties, in which many precursor concepts and pictures are introduced, can be found in Refs. 30-32. And a review on Raman scattering in Fe superconductors has been made in Ref. 33. In this paper, we summarize the existing experimental and theoretical results in both cuprate and iron-based superconductors from the point of view of Raman scattering, and discuss the connection between EPC and HTSC paring mechanism. We will give a short introduction to crystal structures and Raman-active modes of relevant compounds. Then we will discuss EPC in both HTSC systems through Fano asymmetric lineshapes and the renormalization of self-energies crossing SC and other electronic transitions.

**II. Crystal structures and Raman phonon modes in cuprate and Fe-based superconductors**

Generally cuprate superconductors are classified and named according to the heavier elements, for example, rare-earth-, Y-, Bi-, Tl-, Hg-systems, etc. All the compounds contain one or more CuO planes and charge reservoirs. In the following we will take $YBa_2Cu_3O_{7-x}$ (YBCO) as an example to illustrate Raman-active modes in cuprates. As shown in Fig. 1, two buckled CuO planes, charge reservoirs and CuO chains reside on the opposite sides of yttrium symmetrically. Compared to the tetragonal phase ($D_{4h}$) at x=1, the orthorhombic one ($D_{2h}$) at x=0 has a difference of about 1% between a and b lattice parameters and can be seen as a pseudo-tetragonal structure. So, we adopt $D_{4h}$ point group for both phases in symmetry analysis. Raman-active modes are $4A_{1g}+B_{1g}+5E_g$ for $D_{4h}$. In $D_{2h}$ case, $A_{1g}$ and $B_{1g}$ are combined into $A_g$ modes and doubly degenerate $E_g$ is split into $B_{2g}$ and $B_{3g}$ modes. The vibration patterns of several Raman modes in YBCO are illustrated in Fig.2.

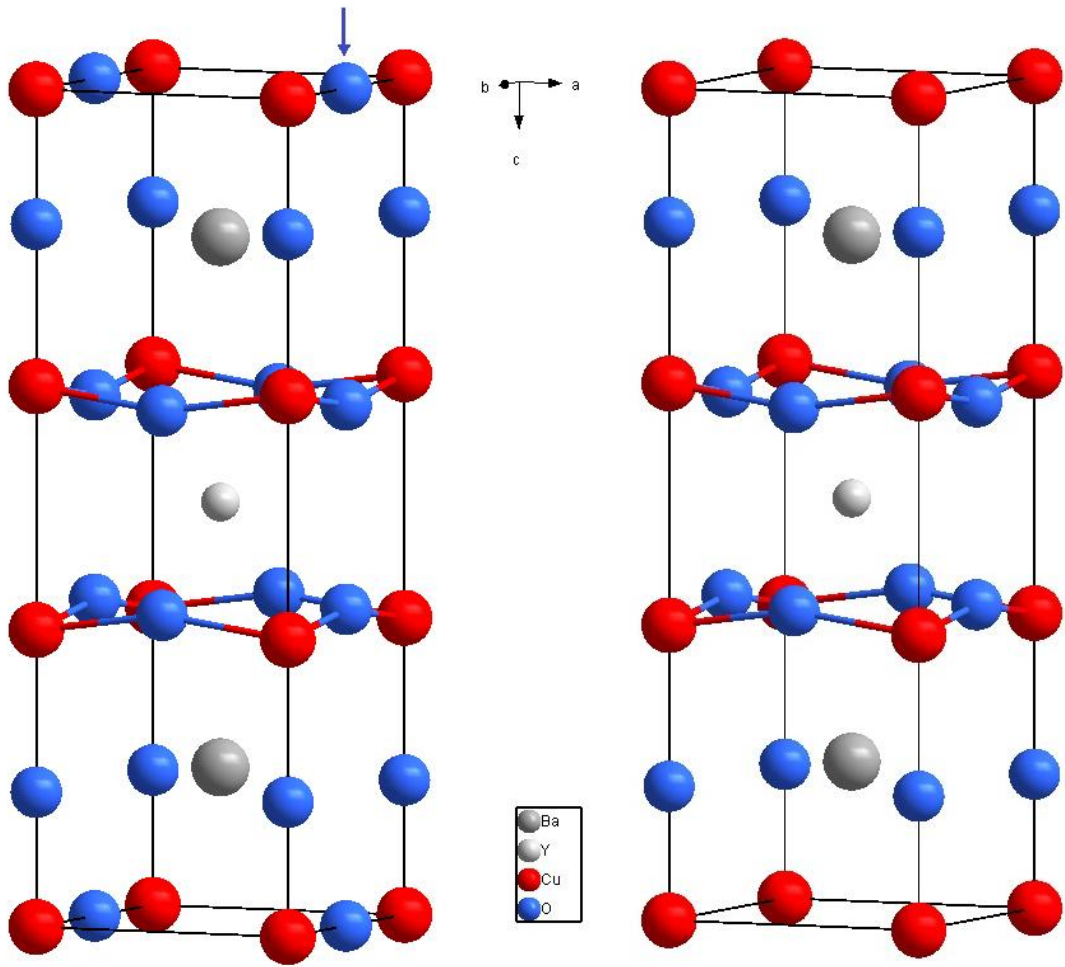

Fig. 1 Crystal structures of YBCO-123. The left is orthorhombic phase and the arrow indicates oxygen atom in CuO chain. The right is tetragonal phase without CuO chain.

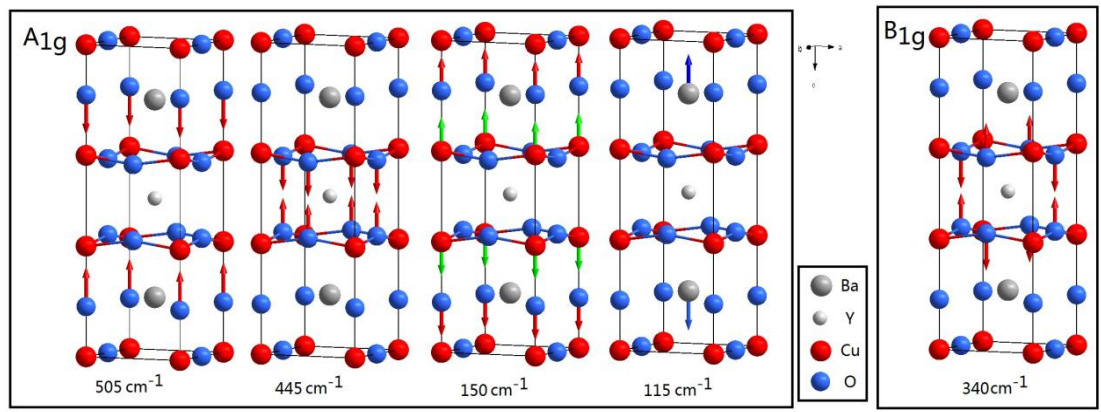

Fig. 2 Five Raman-active modes vibrating along c-axis

Up to now, tens of Fe-based superconductors have been synthesized, basically which can be divided into six categories: 1) "1111", represented by F-doped LaFeAsO, is the first reported iron-based superconducting system. The highest Tc in the series can be raised to 56 K by rare-earth substitution [34]. 2) "122", represented by Co or K-doped $BaFe_2As_2$, is the most studied system due to high-quality crystals available and tunable doping levels [35]. 3) "111" refers to LiFeAs or NaFeAs with a maximum Tc ~ 18 K [36]. 4) "11" refers to Fe(Se,Te), the prototype of iron-based superconductor without poisonous element As [37]. 5) "21311" is isostructural to 122 system with a divalent group instead of Ba [38]. 6) Intercalated FeSe superconductor with a maximum Tc of 30 K, represented by $A_{0.8}Fe_{1.6}Fe_2$ (A = K, Tl, Cs . . . ), was discovered recently [39]. It is the Fe-deficient version of the 122 structure and exhibits a strong Fe-vacancy ordering. Similar to the structure of cuprates, the Fe-based compounds also have conducting FeAs(Se) layers and charge reservoirs. Unlike CuO planes, Fe ions are not located exactly in the same plane as As(Se) ions, but are rather sandwiched by As(Se) ions. Four Raman-active modes of $BaFe_2As_2$ are shown in Fig. 3. Ba, Fe and As ions occupy 2a, 4d and 4e positions, respectively. Ba ions allow a $B_{1g}$ mode vibrating along c-axis and an in-plane $E_g$ mode, and As ions allow an out-of-plane $A_{1g}$ mode and an in-plane $E_g$ mode. Similar Raman modes related to FeAs(Se) exist in the other Fe-based systems. But in 111 and 1111 systems, additional out-of-plane $A_{1g}/B_{1g}$ and $E_g$ modes may be observed because Li/Na, La and O in charge reservoirs are off symmetry center. More details can be found in the Ref. 33.

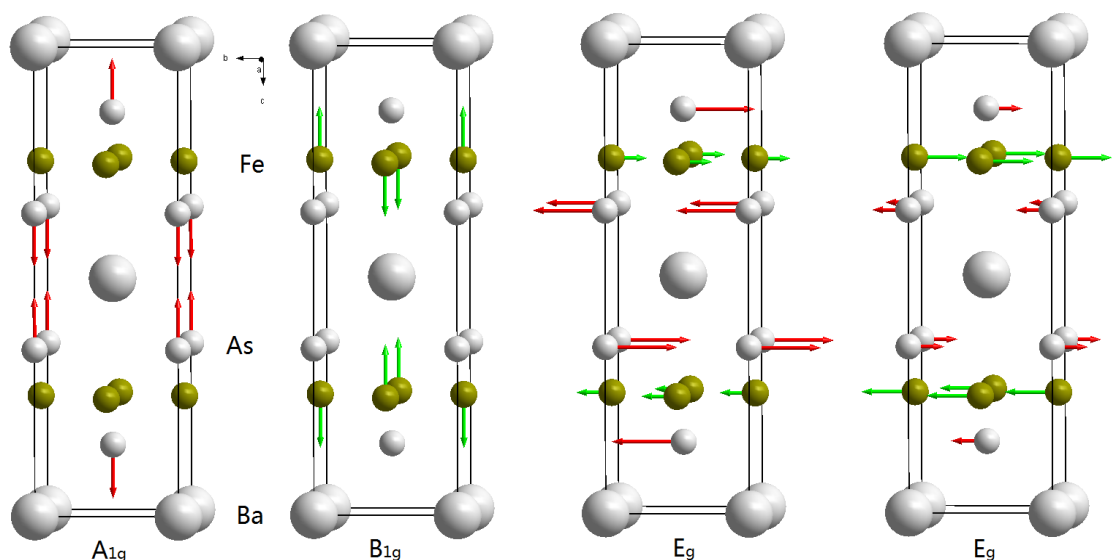

Fig. 3 Crystal structure of $BaFe_2As_2$ and its Raman modes

**III. Electron-phonon coupling in cuprate superconductors**

III.1 Fano effect

The Fano effect was first proposed to explain atomic absorption spectrum [40]. Then it was employed to understand the asymmetric phonon lineshape in heavily doped Si [41, 42], which is caused by the resonance between discrete phonon spectra and the electronic continuum. In Raman spectra, the Fano effect typically changes a Lorentzian phonon peak to an asymmetric one. If there is no interaction between phononic and electronic excitations, one can expect a simple superposition of both contributions. But if there exists a resonance between them, lineshapes and widths will be modified. The scattering intensity modified by EPC is written as [43]:

$$I(\omega) = I_C \frac{|q+\varepsilon|^2}{1+\varepsilon^2} + I_b(\omega) \qquad (1)$$

where $\varepsilon = \omega - \omega_p/\Gamma_p$ and $\omega_p = \omega_p^0 + \delta\omega_p$, $\omega_p^0$ is the bare phonon frequency, $q$ is a parameter quantitatively describing asymmetry, $\Gamma_p$ is the phonon width (half width at half maximum), $I_C$ is an intensity parameter and $I_b(\omega)$ represents incoherent background. An asymmetric peak will get back to Lorentzian lineshape at $q \to \infty$. If neglecting other renormalization mechanism, one can further establish the relation between scattering intensity and microscopic parameters:

$$I(\omega) \sim \pi T_e^2 \rho(\omega) \frac{\left(\omega - \omega_p^0 + T_p V/T_e\right)^2}{\left[\omega - \omega_p^0 - V^2 R(\omega)\right]^2 + [\pi V^2 \rho(\omega)]^2} \qquad (2)$$

where $T_e$ and $T_p$ electronic and phononic Raman vertexes, $V$ electron-phonon coupling vertex, $\omega_p^0$ bare phonon frequency, $\rho(\omega)$ and $R(\omega)$ the imaginary and real parts of electronic

polarizability respectively. The comparison between (1) and (2) gives:

$$q = \frac{T_p V/T_e + V^2 R(\omega)}{\pi V^2 \rho(\omega)}$$

$$\varepsilon = \frac{\omega - \omega_p^0 - V^2 R(\omega)}{\pi V^2 \rho(\omega)}$$

Then EPC strength and other parameters can be obtained by fitting phonon peaks with the above expressions.

III.2 Fano effect in cuprates

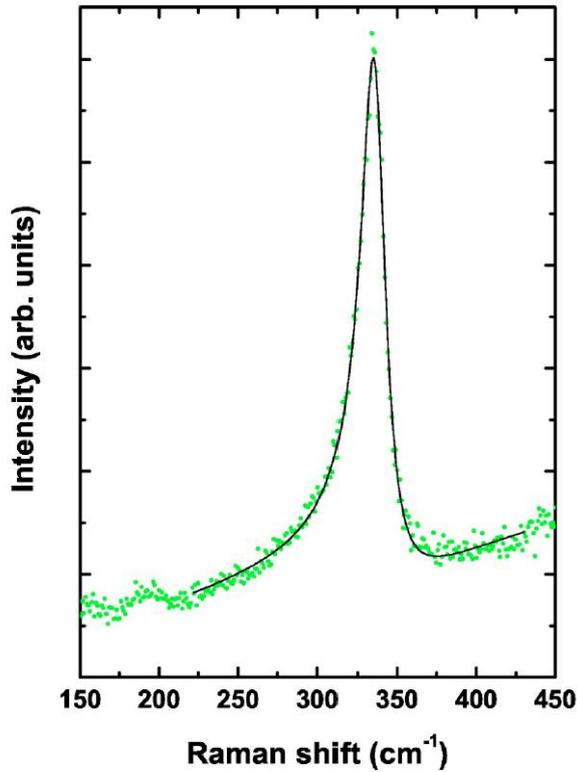

Fig. 4 Fano asymmetric fitting to $B_{1g}$ mode at 340 cm$^{-1}$ in YBCO. The temperature is 16.6 K. The fitting gives ω =336.3±0.1 cm$^{-1}$, q=-6.0±0.2, HWHF γ =9.6±0.1 cm$^{-1}$.[51]

The asymmetry in YBCO has been studied in-depth, especially for the buckling $B_{1g}$ mode of oxygen. Thomsen *et al.* and Cooper *et al.* first reported the asymmetry of the $B_{1g}$ mode at 340 cm$^{-1}$ [44, 45], which was confirmed by further Raman measurements [46–56]. As shown in Fig. 4, the $B_{1g}$ mode exhibits a clear asymmetry, which can be perfectly fitted by a Fano lineshape. The fitting gives quite a large EPC strength (*q* ~-6), which implies that electronic properties may be substantially affected by EPC in YBCO. It seems that EPC in YBCO strongly depends on oxygen content. Actually EPC is much larger in the superconducting YBCO at x ~ 7 and the $B_{1g}$ mode even goes back to a symmetric Lorentzian lineshape in the insulating phase at x ~ 6 [32]. The evolution of asymmetric factor 1/*q* with oxygen content is shown in Fig. 5 [56]. The reduction of asymmetry at small x may be a reflection of low carrier concentration. Consistent with YBCO, no asymmetry was observed in $PrBa_2Cu_3O_7$ which has a very low carrier concentration [57]. However, in $RBa_2Cu_3O_7$ with very close Tc's, different rare-earth R ions produce different q values [48]. This means that the asymmetry is also related to other factors, not only carrier concentration. It was reported that q even depends on polarization geometry [58]. Fano resonance was also reported for the mode at 112 cm$^{-1}$ in YBCO, which was assigned as the vibration of Ba along c-axis (see Fig. 2). But strangely its asymmetry remains unchanged with different rare-earth ions and carrier concentrations [59]. Hence the mode was considered irrelevant to superconductivity.

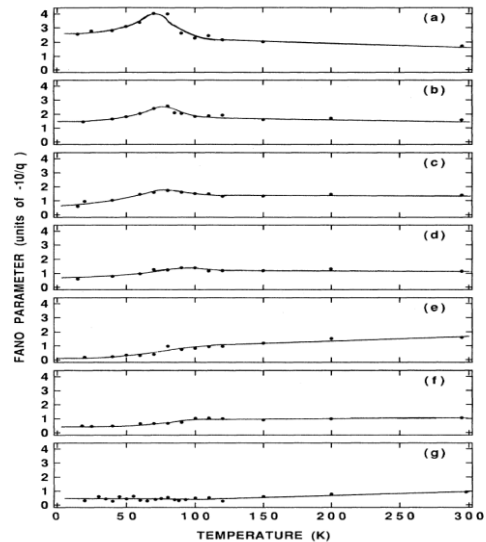

Fig. 5 Temperature dependence of asymmetric factor extracted from B1g phonons. The oxygen contents are 7, 6.99, 6.93, 6 .9, 6.86, 6.81, 6.68 from (a) to (g).[56]

Several scenarios have been proposed to understand the origin of Fano lineshape in YBCO [60-62]. Opel et al. suggested that EPC for the 340 cm$^{-1}$ $B_{1g}$ mode is associated with the buckling of CuO planes [63, 64]. As shown in Fig. 6, $Y^{3+}$ and $Ba^{2+}$ ions are located on the two sides of CuO planes, which gives rise to an effective electric field vertical to CuO planes. The field causes the buckling of CuO planes and a charge transfer between neighboring oxygen in an out-of-phase vibration of oxygen like 340 cm$^{-1}$ $B_{1g}$ mode. In this picture, $B_{1g}$ phonons couple with the transferred electrons and Fano lineshape comes out. EPC strength obtained by fitting $B_{1g}$ mode coincides with that estimated by vertical electric field, which is calculated using the experimental buckling parameters [65, 66]. Generally carrier doping determines the distance between Fermi level and van Hove singularity, and hence affect EPC. This may explain why the asymmetry changes with oxygen content as demonstrated in Fig. 5. There is no such electric field and buckling of CuO planes due to local charge imbalance in Bi-2212 system. By doping Y into Bi-2212 to induce charge imbalance, Opel. et al. found that EPC strength is increased by almost one order of magnitude [63]. At the same time, Tc goes down from 91.5 K to 57.0 K. This seems to imply that EPC is not in favor of superconductivity, which appears to be supported by some theoretical scenarios [67].

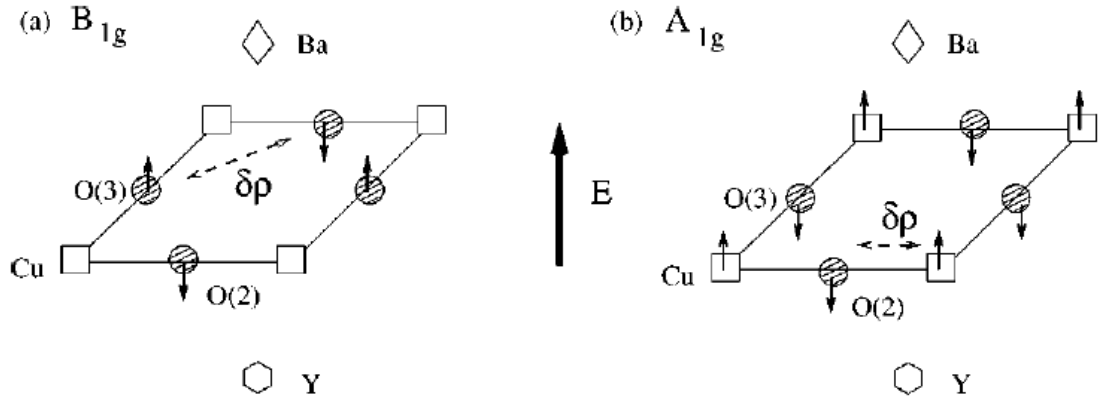

Fig. 6 $B_{1g}$ and $A_{1g}$ modes and the electric field induced by $Ba^{2+}$ and $Y^{3+}$ in YBCO. The eletric field causes a charge transfer between neighboring oxygen.[63]

III.3 Renormalization of self-energy induced by superconducting transition

III.3.1 Phonon frequencies and widths

As discussed above, the imaginary and real parts of electronic self-energy and polarizability can be extracted from phonon lineshapes, through which we can also obtain electronic properties across an electronic transition, for example a SC transition.

In fact, phononic self-energy effect has been extensively investigated in YBCO. Phonon softening at the SC transition was first reported in polycrystals [68], and then observed in single crystals [43]. And it was further confirmed by other techniques [45, 69, 70]. Phonon hardening was also found at the SC transition in A1g mode [71], an in-phase vibration of oxygen in CuO planes along c-axis. The softening and hardening were believed to intimately connect with superconductivity [32]. The reasons are: 1) a structural change which can bring about a phonon frequency shift of 2%, was actually not observed [72, 73]. 2) The phonon anomaly doesn't appear in the non-SC YBa2Cu3O6 and PrBa2Cu3O7 [57, 74]. 3) The temperature Ts at which the phonon anomaly occurs, is found to associate with Tc through the relation *dTs/dH = dTc/dHc2*. It means that phonon anomaly will disappear if superconductivity is killed by magnetic field[75]. Klein and Dierker presented a qualitative understanding based on the BCS theory [76]. And Zeyher and Zwicknagl explained the phonon anomaly in the framework of the Eliashberg theory[77,78]. The basic idea is that EPC will induce a renormalization of phononic self-energy across an SC transition. And the imaginary and real parts reflect the changes in width and frequency respectively, which can be written as [71]:

$$\Delta\Sigma_\nu/\omega_\nu = \lambda_\nu f(\omega_\nu/2\Delta)$$

where $\Delta\Sigma$ the change of phonon self-energy, $\nu$ denotes phonon modes and $\lambda_\nu$ the corresponding EPC constant, f the function of reduced frequency $\tilde{\omega} \equiv \omega_\nu/2\Delta$ and $\Delta$ the SC gap.

In the BCS weak coupling limit, one obtains:

$$f(\tilde{\omega}) \equiv \begin{cases} -2u/\sin 2u & \text{for} \quad \sin u \equiv \tilde{\omega} < 1 \\ (2v - i\pi)/\sinh 2v & \text{for} \quad \cosh v \equiv \tilde{\omega} < 1 \end{cases}$$

The EPC constant $\lambda_\nu$ contributed by the $\nu$th mode is defined as:

$$\lambda_\nu \equiv 2N(0)\langle |g_{\nu kk}|^2\rangle_{FS}/\omega_\nu$$

where N(0) is the electronic density of states and g are EPC matrix elements. For a phonon with $\omega_\nu > 2\Delta$, its frequency will be shifted up and the width broadened, while a phonon with $\omega_\nu < 2\Delta$ will behave exactly in the opposite way. For a d-wave symmetry, the calculations in the clean limit were different from the above results [79], i.e., a phonon with $\omega_\nu < 2\Delta$ will exhibit a softening rather than a hardening and its width will be broadened. Devereaux calculated the temperature dependence of optical phonon modes at low temperatures [80]. It was found that the EPC strength strongly depends on symmetry of the phonon mode. For $B_{1g}$ mode, the real and imaginary parts of the self-energy reach their maximums at $\omega/2\Delta \approx 1$, while the peak shifts to $\omega/2\Delta \approx 0.5$ for $A_{1g}$ mode. He further pointed out that for a d-wave symmetry below Tc the widths of $B_{1g}$ and $A_{1g}$ modes show a $T^3$ and linear temperature dependence, respectively. In an isotropic s-wave superconductor, both have an exponential decay with temperature. This may be used to estimate the SC gap [71]. The earlier studies on phonon anomaly were summarized in Ref. 32. We introduce some later research work in the following.

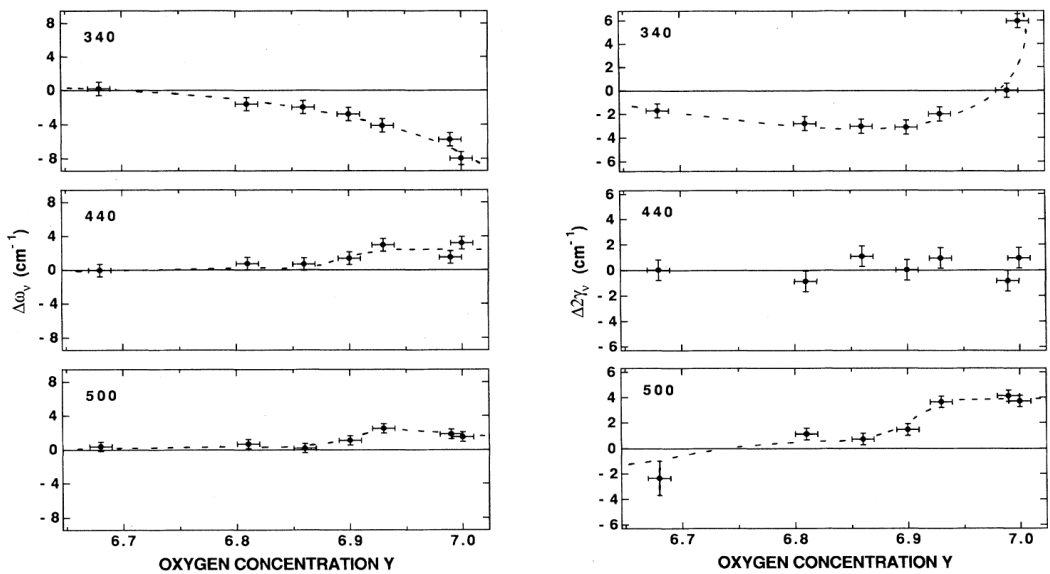

Fig. 7 The evolution of frequencies and widths of Raman phonons at 340, 400 and 500cm$^{-1}$ with oxygen content.[56]

In the early Raman measurements, the changes of phonon frequencies and widths have been already noticed [71, 81], but the magnitude of renormalization seems to be very sensitive to oxygen content [82]. The evolution of phonon frequency and width with temperature in YBCO has been systematically explored [46, 53, 56, 82-84], as shown in Fig. 7. A key question raised from the measurements is whether the renormalization is determined by oxygen content or by carrier concentration. In Ca-doped YBCO-123 and 124 samples, frequency changes induced by SC transitions follow oxygen content but not carrier concentration [52], as shown in Fig. 8. However, Raman measurements in the slightly Ca-doped samples shown in Fig. 9 rather suggests that the frequency changes at Tc is exactly proportional to carrier concentration[85-87].

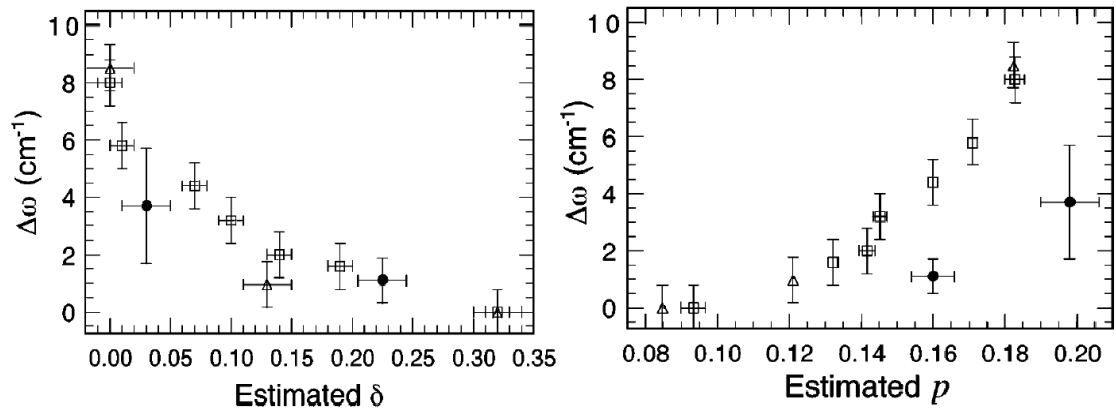

Fig. 8 The frequency softening of B1g mode in Ca doped YBCO in the SC state with oxygen defects (left) and carrier concentration (right).[52]

Generally it is recognized that both phonon self-energy normalization and SC gap are tuned by carrier concentration. The dramatic increase of EPC strength in the overdoped range was considered to be an indication of SC gap size, which was also supported by theories [79]. The reduction of normalization in the underdoping range may be related to the pseudogap [88]. In optimally doped YBCO, Ba $A_{1g}$ mode at 120 cm$^{-1}$ and oxygen $B_{1g}$ mode at 340 cm$^{-1}$ demonstrate renormalization across Tc. In underoped YBCO, phonon anomalies of $B_{1g}$ mode are removed while $A_{1g}$ phonon hardening occurs at 150 K and its width rapidly decreases, which was regarded as a sign of the opening of pseudogap [89].

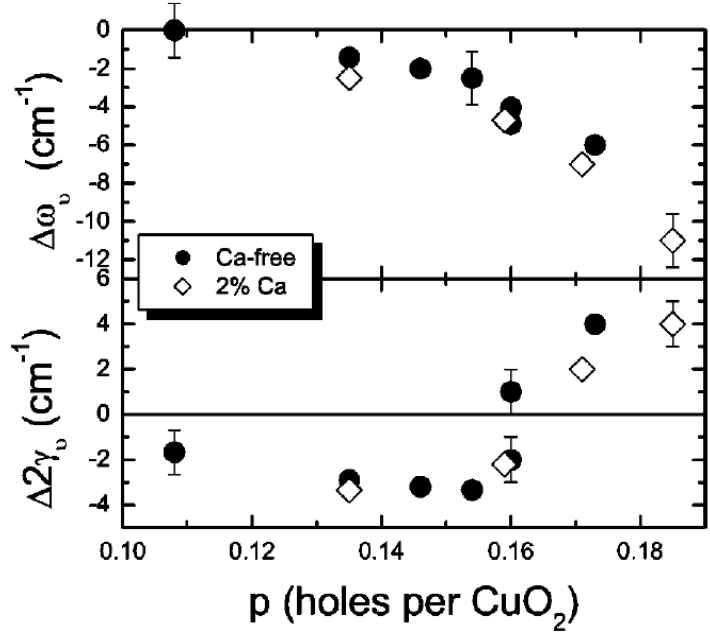

Fig. 9 Carrier concentration dependence of renormalization of $B_{1g}$ phonon in YBCO, where $\Delta\omega \equiv \omega(30\text{ K}) - \omega(100\text{ K})$ $2\Delta\gamma \equiv 2\gamma(T_0) - \Gamma_{AH}(T_0)$.[51]

In oxygen-sufficient YBCO, different lineshapes and peak positions of $B_{1g}$ mode were measured in XX and XY configurations [58]. Assuming constant electronic Raman vertex $T_e$ and phonon Raman vertex $T_p$, Fano fitting gives different frequencies, widths and q. And these parameters indicate anomalies at Tc. This could not be explained by pure d-wave pairing. But it was argued that the assumption on electronic and phononic vertexes is not reasonable [90]. After taking into account both contributions simultaneously, it was concluded that about 10-15% s-wave were mixed into electron pairing [50]. The renormalization of $B_{1g}$ mode was also reported in YBCO films [91] but not observed in NdBaCuO crystals [92].

Besides the $B_{1g}$ mode at 340 cm$^{-1}$, much attention was also paid to the Ba $A_{1g}$ mode at 120 cm$^{-1}$. In YBCO-124 system, the mode shows a clear softening and broadening, which was associated with the opening of a small gap at 113 cm$^{-1}$ [93]. The anomalies of $A_{1g}$ mode at Tc are strongly suppressed in Ni and Zn-doped YBCO-124. The relation between gap size and the normalized frequencies and widths follows the predictions by Zeyher and Zwicknagl based on s-wave [78]. McCarty et al. reported the anomalies of the modes at 440 and 500 cm$^{-1}$ at Tc, and further estimated the SC gap [94]. Two Raman-inactive modes in underdoped YBCO-123 exhibit a significant resonance and an obvious softening in the SC state [54]. The origin of the strange EPC remains unknown.

The studies of phonon renormalization in Bi-2212 system are relatively dispersive. Burns et al. and Boekholt et al. reported that a softening occurs in $A_{1g}$ mode of O(3) in SrO at 465 cm$^{-1}$ [95,96]. This was also seen by Martin et al. [97,98]. Below Tc Leach et al. observed an anomalous drop of width of $B_{1g}$ mode at 285 cm$^{-1}$ [99], which is exactly the same $B_{1g}$ mode at 340 cm$^{-1}$ in YBCO. But no frequency change is found for $B_{1g}$ mode. These results are still waiting for a consistent understanding.

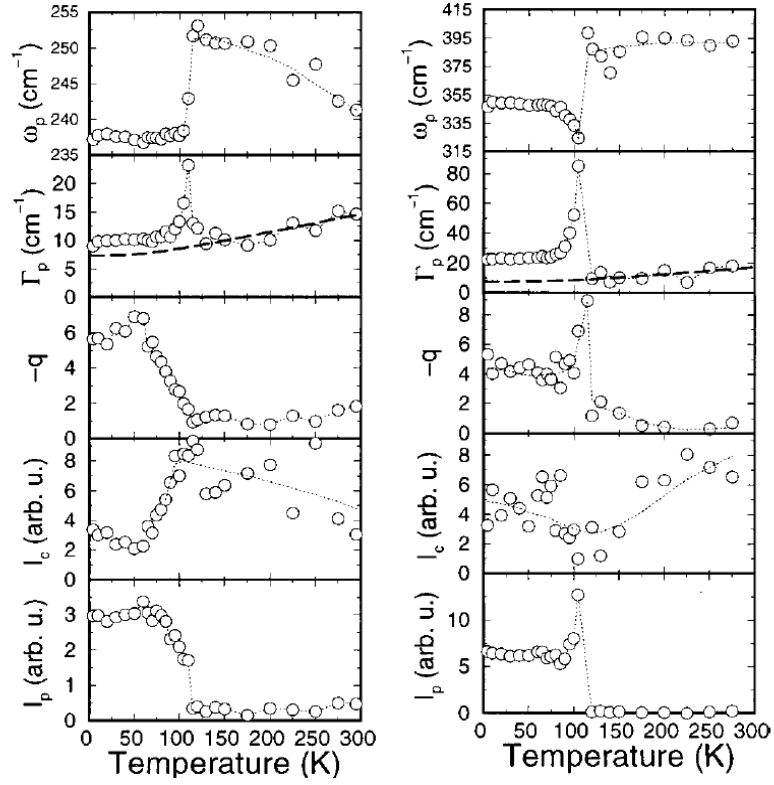

Fig. 10 Fitting results of $A_{1g}$ modes at 240 cm$^{-1}$ (left) and 390 cm$^{-1}$ (right). The dashed lines are the fitting to the data in the normal state.[103]

In Hg-1201 the softening of $A_{1g}$ mode at 568 cm$^{-1}$ below Tc was observed and accompanied by a broadening [100]. Actually there exist phonon anomalies at Tc in Hg-1223 system [101]. Different from YBCO, neither renormalization nor Fano lineshape was seen in $B_{1g}$ mode of oxygen in CuO planes at 245 cm$^{-1}$. This may be due to the flat [102] rather buckled CuO planes [63, 64]. Interestingly, a large asymmetry was found in $A_{1g}$ mode of in-plane oxygen. The EPC observed in Hg-1234 may be the strongest among cuprates, as shown in Fig. 10 [103]. A sharp drop in frequency was probed at Tc in the $A_{1g}$ modes of Ca at 390 cm$^{-1}$ and in-plane oxygen. At the same time their widths are doubled across SC transition. Using Allen formula [104], the EPC strength $\lambda_\nu$ is about 0.08. Suppose that EPC strengths of all the 60 modes are roughly at the same level, we will obtain a total EPC strength $\lambda \sim 6$, which is high enough to explain the high Tc.

All the results mentioned above lead to the conclusion that unambiguous phonon renormalization induced by SC transition can be seen in most cuprate systems. The magnitude of renormalization depends on many factors, such as cuprate systems, phonon frequency and symmetry and doping etc. The most prominent manifestations are in-plane oxygen $B_{1g}$ mode in YBCO-123 and $A_{1g}$ modes in YBCO-124 and Hg-system. Phonon renormalization caused by EPC provides an insight into pairing mechanism.

III.3.2 Phonon intensities

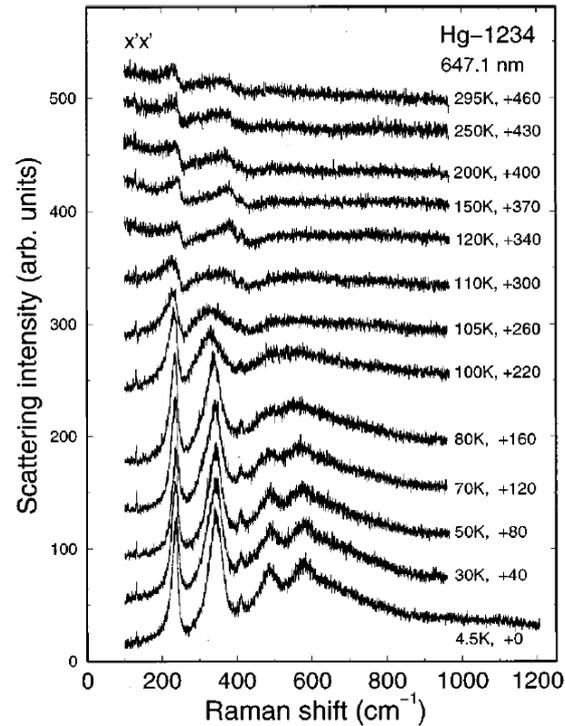

Fig. 11 Raman spectra at various temperatures in Hg-1234 with Tc ~ 123 K.[103]

Besides phonon frequency and width, large changes in phonon intensity were also found in many cuprate superconductors when crossing Tc [29,101,103,105-110]. A representative example is Hg-1234 [103], as shown in Fig. 11. In the normal state, the modes at 240 and 390 cm$^{-1}$ are quite weak and the modes at 487 and 575 cm$^{-1}$ are almost invisible. All the four modes have an $A_{1g}$ symmetry. The modes are dramatically enhanced in the SC state. The enhancement of intensity was considered to be contributed by Fano resonance. In YBCO, the integrated intensities of Ba $A_{1g}$ mode increase several times below Tc, but only in XX channel. In Bi-2212, intensity anomaly is more obvious using an excitation laser of 633 nm than 458 nm. A resonant study of RBCO-123 (R=Yb, Er, Sm, Nd) revealed that O(4) mode in X'X' channel and O(2)-O(3) mode in X'Y' channel become stronger in the SC state when excitation energy is larger than 2.1 eV, while they are suppressed using an excitation laser less than 2.1 eV [110]. The characteristic energy does not depend on carrier concentration. Similar resonance phenomena were also observed in YBCO thin films deposited on SrTiO$_3$ substrate [109]. Sherman et al. proposed an EPC picture to explain the above results [107]. Generally Raman intensity of $B_{1g}$ mode is determined by interband EPC and the coupling strength depends on the ratio of EPC matrix elements to interband gap. The smaller the gap is, the larger the coupling will be. SC transition effectively extends the phase volume of electrons intermediating scattering process, due to a small gap and a large density of states at Fermi level in the SC state. This explains why the scattering intensity is raised across SC transition. The intensity anomaly of $A_{1g}$ mode was analysed in the framework of weak-coupling BCS theory [108]. The excitation energy dependence of O(4) mode intensity mentioned above [110], could be also understood with this picture. The change of force constants induced by SC transition, also seems to be a plausible explanation for intensity anomaly at Tc [55].

**IV. Electron-phonon coupling in iron-based superconductors**

There are not so many Raman studies of EPC in iron-based superconductors as in cuprates. In 1111 system, no phonon asymmetry was reported so far. Phonon anomalies are absent at structural, spin-density-wave (SDW) and SC transitions. It implies that EPC and phonon renormalization are weak in the system [111-113]. Similarly, in 111 system without structural and magnetic phase transitions, the observations are basically close to those in 1111 [114]. While in 11 and 122 systems, Fano linshapes and renormalization were observed. We discuss this in more details in the following.

IV.1 Fano effect in Fe superconductors

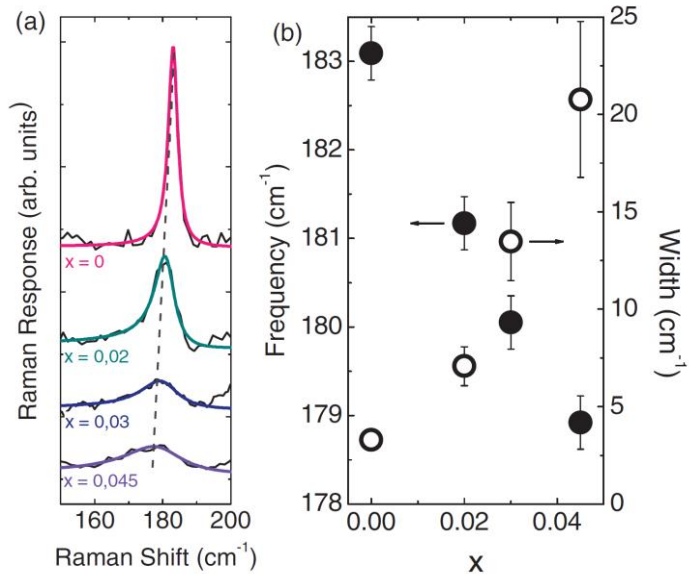

Fig. 12 (a) The evolution of $A_{1g}$ phonons with doping in $Ba(Fe_{1-x}Co_x)_2As_2$. The solid lines are Fano fitting results. (b) Doping dependence of frequencies and widths of $A_{1g}$ phonon.[115]

Fano effect was explicitly reported in Co-doped $BaFe_2As_2$ [115], as shown in Fig. 12. The asymmetry of $A_{1g}$ phonon linshape appears in the SDW state. And the asymmetry becomes more clear with increasing doping, indicating an enhancement of EPC. The changes in frequency and width are too large to be explained using a c-axis change of about 0.3%. In point of view of EPC, the phonon softening and broadening are a consequence of the increase of the density of states at the Fermi level induced by carrier doping. It is quite similar to the case of cuprates, in which phonon asymmetry strongly depends on doping level. The conclusion seems to be supported also by the calculations [116]. Furthermore, it was found that a spin-ordered state also favors EPC. The maximum EPC strength λ even reaches up to 0.35 in 122. But the value is still not large enough to estimate Tc.

IV.2 Renormalization of self-energy induced by SDW and SC transitions

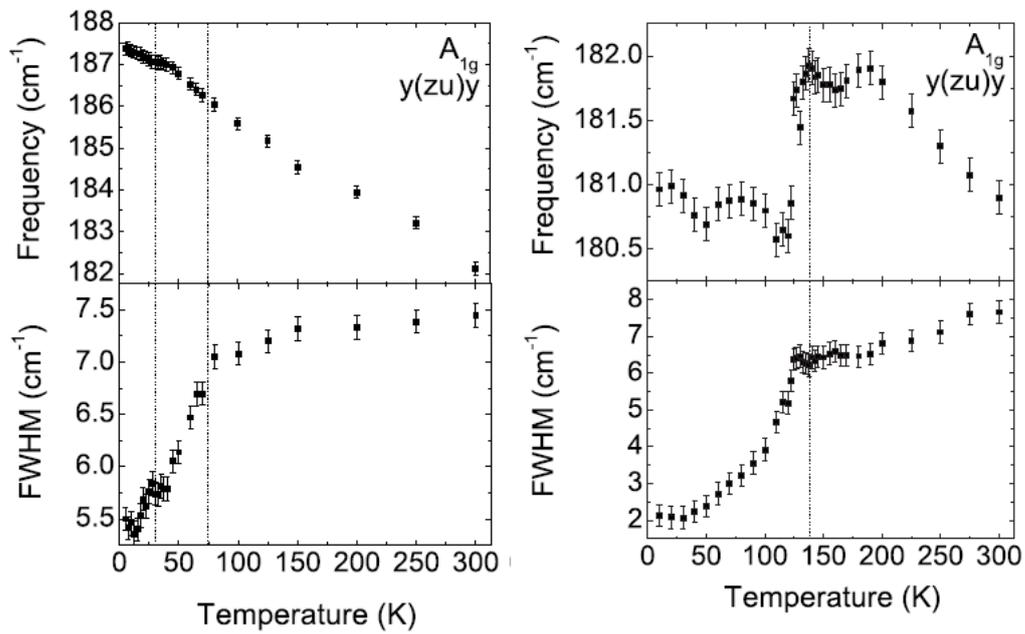

Fig. 13 The evolution of frequencies and widths of $A_{1g}$ mode with temperature in underdoped $Ba_{0.72}K_{0.28}Fe_2As_2$ (Tc=29 K，Ts=75 K) and parent compound $BaFe_2As_2$ (Ts=138 K).[121]

So far there was no consensus on whether or not phonon self-energy renormalization occurs at the structural, SDW and SC transitions in Fe-based compounds. Litvinvhuk *et al.* reported that no anomaly was found in $Sr_{1-x}K_xFe_2As_2$ (x=0, 0.4) [117], similar to the results in 1111. While in $CaFe_2As_2$, a new $A_{1g}$ mode appears below SDW transition temperature and $B_{1g}$ mode exhibits an anomalous broadening and a hardening of ~4 $cm^{-1}$ [118]. It can be well understood in term of phonon renormalization due to SDW gap rather than a simple lattice change. The SDW gap was probed in optical absorption spectra [119]. Phonon renormalization caused by SDW gap was also in parent compound $BaFe_2As_2$ [120] and hole-doped $R_{1-x}K_xFe_2As_2$ (R=Ba, Sr) [121]. Fig. 13 shows the behavior of $A_{1g}$ mode in $Ba_{0.72}K_{0.28}Fe_2As_2$ and $BaFe_2As_2$ at SDW transition temperature. Compared to the change in parent compound, $A_{1g}$ renormalization effect is weaker in K-doped samples. And similar results were obtained but weaker in the $B_{1g}$ channel. This was supported by first-principles calculations, which indicate that EPC related to As ions is stronger [122-124]. Besides, a small softening of 0.3 cm$^{-1}$ and a kink in width appear at Tc, which was also reported in $Sr_{0.85}K_{0.15}Fe_2As_2$ (Tc = 29 K). This was considered to be phonon self-energy renormalization associated with the opening of SC gap, which implies that the SC gap 2Δ should be samller than $A_{1g}$ phonon frequency, .i.e., 2Δ ≦ 23 meV. Actually the renormalization induced by SC gap was also observed in B1g channel in $Pr_{0.12}Ca_{0.88}Fe_2As_2$ [125].

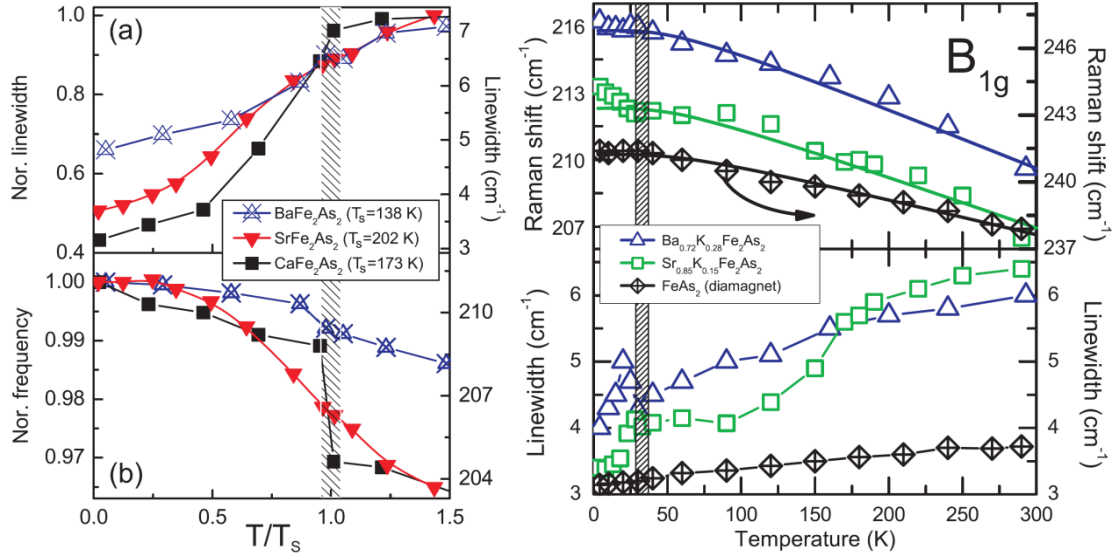

Fig. 14 Left: Temperature dependence of normalized linewidth and frequency in the three 122 parent compounds. Right: Raman frequencies and widths of $B_{1g}$ mode in two parent compounds and reference compound $FeAs_2$.[126]

A systematic study on phonon renormalization in 122 system is shown in Fig. 14 [126]. In the three parent compounds, a clear kink in both frequency and width can be seen at SDW transition temperatures. This was explained as phonon renormalization due to the opening of SDW gap. The larger and discontinuous jump in $CaFe_2As_2$ is characteristic of a first-order structural transition rather than SDW transition. In SC 122 samples, the hardening and softening of $B_{1g}$ mode caused by SC transition could be evaluated after subtracting contributions from SDW transition. The values give us an EPC constant $\lambda_{B1g} \approx 0.02$ using Allen formula $\Gamma = 2\pi \lambda N(0) \omega^2$ [104], where the phonon width $\Gamma$ is related to EPC, ω is the phonon frequency, $\lambda$ is the EPC strength and N(0) is the density of states at Fermi surface. Phonon renormalization provides an alternative way to get EPC strength through the formula $\lambda = -K \operatorname{Re}[(\sin u)/u]$ with $K = \frac{\omega^{sc}}{\omega^N} - 1$ and $u \equiv \pi + 2i \cosh^{-1}(\omega^N/2\Delta)$, where

$\omega^{SC}$ and $\omega^{N}$ are the phonon frequencies in the SC state and normal state, respectively. This outputs $\lambda_{B_{1g}} \approx 0.01$. The EPC strengths estimated by both methods are far smaller than 1, which was required by strong-coupling theories if one would explain the high Tc with a pure EPC picture [122]. This seems to imply that EPC may play a minor role in Fe-based superconductors.

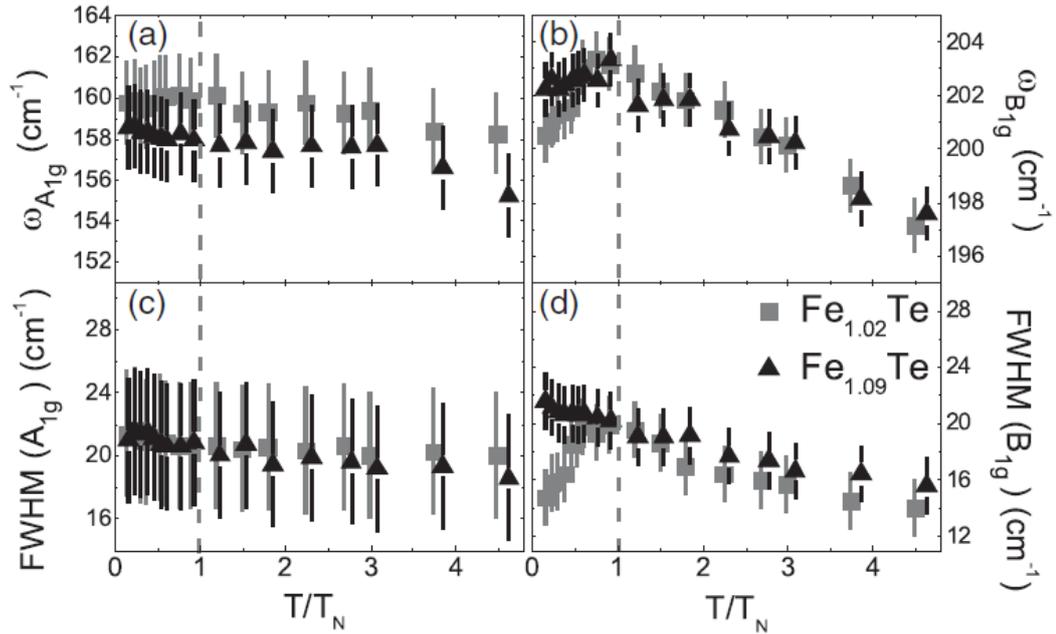

Fig. 15 Temperature dependence of frequencies and widths of $A_{1g}$ and $B_{1g}$ phonons in FeTe parent compounds with different Fe content.[127]

Phonon anomalies across SDW and structural transitions were also reported in the 11 system. The $B_{1g}$ mode in $Fe_{1.02}$Te shows an anomalous kink in frequency near the transition temperatures [127, 128]. With the substitution of Te by Se, both SDW transition and the kink are suppressed, which suggests that the kink is a reflection of the opening of SDW gap (Fig. 15). Actually the suppression was also found when doping excess Fe into parent compounds [129, 130]. Similarly, phonon renormalization is weak in 11 compounds with excess Fe. Phonon anomalies caused by the SDW transition indicate the existence of EPC in 11 system.

EPC was observed in the newly discovered $K_xFe_{2-y}Se_2$ system. Raman measurements by Zhang et al. revealed that the Ag mode at 180 cm$^{-1}$ exhibits a hardening of ~1.5 cm$^{-1}$ when entering into the SC state [131], as shown in Fig. 16. The indication of renormalization suggests a SC gap $2\Delta \leqq 22.5 meV$ (180 cm$^{-1}$), which is close to the SC gap of ~20 meV measured by angle-resolved photoemission [132]. In addition, the rapid decrease of integrated intensities of Se Ag mode at 66 cm$^{-1}$ with temperature, is similar to those of $A_{1g}$ modes in cuprate superconductors (Fig. 11), which was considered as evidence for EPC.

So far, we find two kinds of renormalization phenomena in Fe-based superconductors. One is associated with the opening of SDW gap, as displayed in 11 and 122 systems. This is evidence for EPC and further calculations demonstrated that EPC favors a spin-order state. And there exist some experimental indications of spin-lattice coupling in the compounds. This implies a tight binding of spin, lattice and charge degrees of freedom in the two systems. Though EPC may play a minor role in pairing, it will have an important impact on superconductivity [115, 116]. The other one is the renormalization induced by the SC transition. Up to now, the effect was reported only in 122 and $K_{0.8}Fe_{1.6}Se_2$ systems. The reason why it is weak in other Fe-based systems may be that the mismatch between SC gap and relevant phonons is relatively large. As mentioned above, phonon renormalization gives $\lambda_{B1g} \approx 0.02$[126]. As comparison, EPC strengths of 0.2 and 0.12 were estimated by femtosecond time-resolved photoemission [133] and pump-probe optical reflectivity [134], respectively. First-principles calculations produce $\lambda \sim 0.23$ [123], 0.35 [116], 0.37 [135]. Raman scattering gives a smaller EPC strength because it measures partial phonon branches at the Brillouin zone center. Even using the calculated EPC strengths, one cannot satisfactorily explain the Tcs.

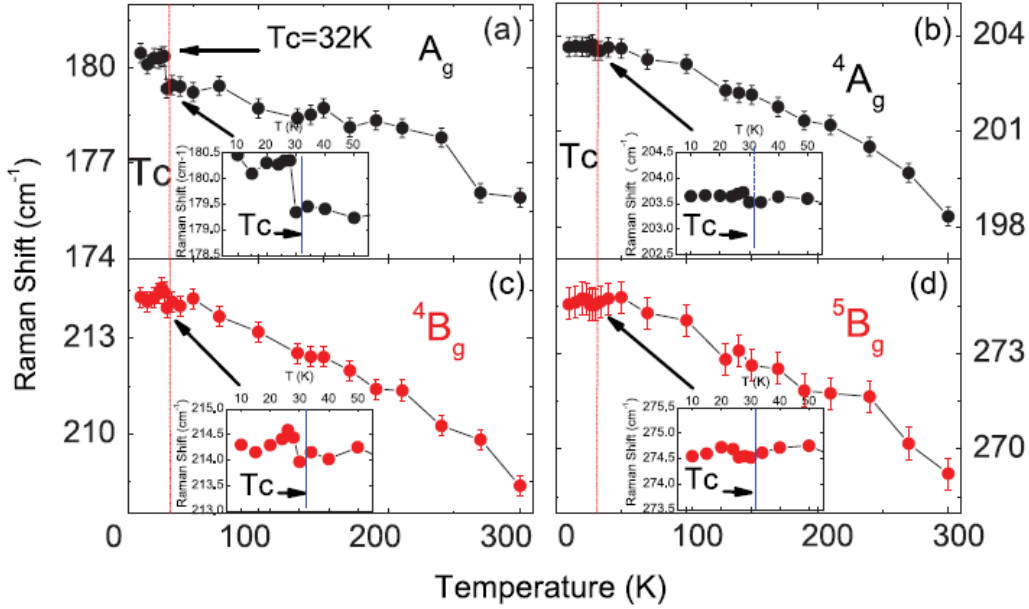

Fig. 16 Temperature dependence of Raman frequencies of selected modes in $K_{0.8}Fe_{1.6}Se_2$.[131]

The experimental and calculated EPC strengths in Fe-based systems are summarized in Table I. As comparison, EPC strengths of cuprate, $MgB_2$ and A15 $Nb_3Sn$ superconductors are also listed in the table. In 11 system, the experimental EPC strength is $\sim 0.16$ [137, 138], and the calculated values are 0.17 [138], 0.15-0.39 [139] and 0.3 [140]. Like the case in 122, the strengths are too small to explain Tcs. There is no consensus on EPC in 1111. The EPC strengths were estimated to 0.25 and

1.53 by resistivity in non-SC and SC PrFeAsO$_{1-x}$F$_x$ polycrystals [141], which seems to support a BCS explanation on Tc ~ 50 K. However, optical pump-probe femtosecond spectroscopy measured λ ~ 0.18 [142] and first-principles calculations give λ ~ 0.21 in 1111[122], which predicts a Tc of ~ 0.8 K. Further exploration of EPC in 1111 system is still waiting for high-quality single crystals available. The similar case is found in 111 system. There is a big difference between EPC strengths presented by ARPES and band calculations, which are 1.53 and 0.29 respectively.

Table I Experimental and calculated EPC strengths in Fe-based and other superconductors

| System | Experimental λ | Calculated λ |
|---|---|---|
| 11 | 0.16-0.01[136], 0.16[137] | 0.17[138], 0.15-0.39[139], 0.3[140] |
| 111 | 1.38[143] | 0.29[144] |
| 1111 | 0.25(NSC)[141], 1.53(SC)[141], 0.18[142] | 0.21[122] |
| 122 | 0.02[126], 0.2[133], 0.12[134] | 0.23[123], 0.35[116], 0.37[135] |
| 122* | / | 0.19-0.34[139] |
| YBCO-123 | 0.016-0.020[47] | 0.27[148] |
| MgB$_2$ | 0.3[158] | 1.01[159] |
| Nb$_3$Sn | 1.80[160] | 1.19[161] |

**V. Conclusions**

The consequences of EPC observed in cuprate superconductors, such as asymmetric phonon lineshape, softening or hardening, broadening, intensity changes etc, are exactly demonstrated in Fe-based systems. The weaker EPC in Fe-based systems may be due to stronger screening effect. The measured EPC strength is roughly in agreement with the calculated one in Fe-based systems, but it seems not to be a dominant factor for pairing. On the other hand, due to the strong coupling among electrons, spins and phonons, EPC effect on superconductivity remains to be revealed.

The situation is more complicated in cuprate superconductors. A lot of calculations demonstrated that EPC alone is not enough to explain so high Tc [145-148]. Even some calculations [149, 150] indicated that EPC is against pairing. On the experimental side, both ARPES [151] and STM [152] suggested that EPC is a key factor for pairing, which is supported by some theoretical calculations [153, 154]. Egami [155] proposed that EPC in cuprates is a more complicated spin-charge-phonon coupling rather than BCS-like, due to spin-dependent charge transfer mediated by phonons in strongly correlated systems. And a d-wave paring can be realized in EPC system with the aid of spin fluctuations [156]. Kulic considered that EPC is just the paring glue in cuprates after summarizing many experiments [157]. So, though a clear microscopic picture is absent, EPC is substantially involved in pairing in some way in both high-Tc superconductors. Considering their parent compounds are spin-ordered, it allows us to conclude that a delicate coupling between electrons, spins and phonons may be crucial to high-Tc superconductivity.

**Acknowledgements**

This work was supported by the NSF of China, the 973 program (Grant Nos. 2011CBA00112 and 2012CB921701).

**References**

[1] Onnes H K 1911 Leiden Commun. 28 120

[2] Gorter C J and Casimir H B G 1934 Physica 1 306

[3] London F and London H 1935 London: Proc. Roy. Soc. 149 71

[4] Bardeen J, Cooper L N and Schrieffer J R 1957 Phys. Rev. 108 1175

[5] Bednorz J D and M üller K A Z 1986 Phys. B 64 189

[6] Wu M K, Ashburn J R, Torng C J, Hor P H, Meng R L, Gao L, Huang Z J, WangY Q and Chu C W 1987 Phys. Rev. Lett. 58 908

[7] Zhao Z X, Chen L Q, Yang Y S, Huang Y Z, Chen G H, Tang R M, Liu Z R, Cui C G, Chen L, Wang L Z, Guo S Q, Li S L and Bi J Q 1987 Chin. Sci. Bull. 32 414

[8] Ginsberg D M 1994 Physical Property of High Temperature Superconductor (vol. 4) (Singapore: World Scientific) p. 3

[9] Batlogg B, Cava R J, Jayaraman A, van Dover R B, Kourouklis G A, Sunshine S, Murphy D W, Rupp L W, Chen H S, White A, Short T K, Mujsce A M and Rietman E A 1987 Phys. Rev. Lett. 58 2333

[10] Leary K J, zur Loye H C, Keller SW, Faltens T A, HamWK, Michaels J N and Stacy A M 1987 Phys. Rev. Lett. 59 1236

[11] Thomsen C, Mattausch H, Bauer M, Bauhofer W, Lui R, Genzel L and Cardona M 1988 Solid State Commun. 67 1069

[12] Katayama-Yoshida H, Hirooka T, Oyamada A, Okabe Y, Takahashi T, Sasaki T, Ochini A, Suzuki T, Mascarenhas A J, Pankove J I, Ciszek T F, DebS K, Goldfarb R B and Li Y 1988 Physica C 156 481

[13] Koch R H, Umbach C P, Clark G J, Chaudhari P and Laibowitz R B 1987 Appl. Phys. Lett. 51 200

[14] Gough C E, Cochough M S, Morgen E, Jordan R G, Keene M, Muirhead C M, Rae A I M, Thomas N, Abell J S and Sutton S 1987 Nature 326 855

[15] Niemeyer J, Dietrich M R and Politis C 1987 Z. Phys. B 67 155

[16] Parmenter R H 1987 Phys. Rev. Lett. 59 923

[17] Varma C M, Schmitt-Rink S and Abrahams E 1987 Solid State Commun. 62 681

[18] Kamar ás K, Porter C D, Doss M G, Herr S U, Tanner D B, Bonn D A, Greedan J E, O'Reilly A H, Stager C V and Timusk T 1987 Phys. Rev. Lett. 59 919

[19] Alexandrov A S 1989 Physica C 158 337

[20] Prelovsek P, Rice T M and Zhang F C 1987 J. Phys. C 20 L229

[21] Emin D 1989 Phys. Rev. Lett. 62 337

[22] Bedell K S and Pines D 1988 Phys. Rev. B 37 3730

[23] Scalapino D J, Loh E and Hirsch J 1986 Phys. Rev. B 34 8190

[24] Pintschovius L and Reichardt W 1998 Physics and Chemistry of Materials with Low-Dimensional Structures (vol. 20) (Dordrecht: Kluwer Academic) p. 165

[25] Pintschovius L 2005 Phys. Stat. Sol. B 242 30

[26] Tohyama T and Maekawa S 2000 Supercond. Sci. Technol. 13 R17

[27] Fink J, Borisenko S, Kordyuk A, Koitzsch A, Geck J, Zabolotnyy V, Knupfer M, B üchner B and Berger H 2007 Lect. Notes Phys. 715 295

[28] Yoshida T, Zhou X J, Lu D H, Komiya S, Ando Y, Eisaki H, Kakeshita T, Uchida S, Hussain Z, Shen Z X and Fujimori A 2007 J. Phys.: Condens. Matter 19 125209

[29] Zhou X J, Cuk T, Devereaux T, Nagaosa N and Shen Z X 2007 Handbook of High-Temperature Superconductivity: Theory

and Experiment (1st edn.) (New York: Springer) pp. 87–144

[30] Thomsen C and Cardona M 1989 High-Temperature Superconductors I and II (1st edn.) (Singapore: World Scientific) p. 409

[31] Devereaux T P and Hackl R 2007 Rev. Mod. Phys. 79 175

[32] Thomsen C 1991 Light Scattering in Solids VI (1st edn.) (Heidelberg: Springer) p. 285

[33] Zhang A M and Zhang Q M 2012 Mod. Phys. Lett. B 26 1230020

[34] Kamihara Y, Watanabe T, Hirano M and Hosono H 2008 J. Am. Chem. Soc. 130 3296

[35] Rotter M, Tegel M and Johrendt D 2008 Phys. Rev. Lett. 101 107006

[36] Wang X C, Liua Q Q, Lva Y X, Gao W B, Yanga L X, Yua R C, Lia F Y and Jin C Q 2008 Solid State Commun. 148 538

[37] Hsu F C, Luo J Y, Yeh K W, Chen T K, Huang T W, Wu P M, Lee Y C, Huang Y L, Chu Y Y, Yan D C andWu M K 2008 Proc. Natl. Acad. Sci. 105 14262

[38] Zhu X Y, Han F, Mu G, Cheng P, Shen B, Zeng B and Wen H H 2009 Phys. Rev. B 79 220512

[39] Guo J G, Jin S F, Wang G, Wang S C, Zhu K X, Zhou T T, He M and Chen X L 2010 Phys. Rev. 82 180520

[40] Fano U 1961 Phys. Rev. 124 1866

[41] Cerdeira F, Fjedly T A and Cardona M 1973 Solid State Commun. 13 325

[42] Cerdeira F, Fjedly T A and Cardona M 1973 Phys. Rev. B 8 4734

[43] Abstreiter G, CardonaMand Pinczuk A 1984 Light Scattering in Solids IV (1st edn.) (Berlin: Springer) p. 128

[44] Thomsen C, Cardona M, Gegenheimer B, Liu R and Simon A 1988 Phys. Rev. B 37 9860

[45] Cooper S L, Klein M V, Pazol B G, Rice J P and Ginsberg D M 1988 Phys. Rev. B 37 5920

[46] Altendorf E, Chrzanowski J, Irwin J C, O'Reilly A and Hardy W N 1991 Physica C 175 47

[47] McCarty K F, Radousky H B, Liu J Z and Shelton R N 1991 Phys. Rev. B 43 13751

[48] Friedl B, Thomsen C and Cardona M 1990 Phys. Rev. Lett. 65 915

[49] Krantz M, Rosen H J, Macfarlane R M and Lee V Y 1988 Phys. Rev. B 38 4992

[50] Bakr M, Schnyder A P, Klam L, Manske D, Lin C T, Keimer B, Cardona M and Ulrich C 2009 Phys. Rev. B 80 064505

[51] Hewitt K C, Chen X K, Roch C, Chrzanowski J, Irwin J C, Altendorf E H, Liang R, Bonn D and Hardy W N 2004 Phys. Rev. B 69 064514

[52] Quilty J W and Trodahl H J 2000 Phys. Rev. B 61 4238

[53] Altendorf E, Irwin J C, Liang R and Hardy W N 1992 Phys. Rev. B 45 7551

[54] Panfilov A G, Limonov M F, Rykov A I, Tajima S and Yamanaka A 1998 Phys. Rev. B 57 R5634

[55] Misochko O V, Sherman E Y, Umesaki N, Sakai K and Nakashima S 1999 Phys. Rev. B 59 11495

[56] Altendorf E, Chen X K, Irwin J C, Liang R and HardyWN 1993 Phys. Rev. B 47 8140

[57] Thomsen C, Liu R, Cardona M, Amador U and Morán E 1988 Solid State Commun. 67 271

[58] Limonov M F, Rykov A I, Tajima S and Yamanaka A 1998 Phys. Rev. Lett. 80 825

[59] Bogachev G, Abrashev M, Iliev M, Poulakis N, Liarokapis E, Mitros C, Koufoudakis A and Psyharis V 1994 Phys. Rev. B 49 12151

[60] Virosztek A and Ruvalds J 1992 Phys. Rev. B 45 347

[61] Zawadowski A and Cardona M 1990 Phys. Rev. B 42 10732

[62] Devereaux T P and Kampf A P 1999 Phys. Rev. B 59 6411

[63] Opel M, Hackl R, Devereaux T P, Virosztek A, Zawadowski A, Erb A, Walker E, Berger H and Forro L 1999 Phys. Rev. B 60 9836

[64] Devereaux T P, Virosztek A and Zawadowski A 1995 Phys. Rev. B 51 505

[65] Jorgensen J D, Beno M A, Hinks D G, Soderholm L, Volin K J, Hitterman R L, Grace J D, SohullerI K, Segre C V, Zhang K and Kleefisch M S 1987 Phys. Rev. B 36 3608

[66] Kaldis E, Röhler J, Liarokapis E, Poulakis N, Conder K and Loeffen P W 1997 Phys. Rev. Lett. 79 4894

[67] Savrasov S Y and Andersen O K 1996 Phys. Rev. Lett. 77 4430

[68] Macfarlane R M, Rosen H J and Seki H 1987 Solid State Commun. 63 831

[69] Feile R, Schmitt U, Leiderer P, Schubert J and Poppe U 1988 Z. Phys. B 72 141

[70] Wittlin A, Liu R, Cardona M, Genzel L, König W, Bauhofer W, Mattausch H, Simon A and Garcia-Alvarado F 1987 Solid State Commun. 64 477

[71] Thomsen C, Cardona M, Friedl B, Rodriguez C O, Mazin I I and Andersen O K 1990 Solid State Commun. 75 219

[72] Jorgensen J D, Veal B W, Paulikas A P, Nowicki L J, Crabtree G W, Claus H and Kwok W K 1990 Phys. Rev. B 41 1863

[73] Horn P M, Keane D T, Held G A, Jordon-Sweet J L, Kaiser D L, Holtzberg F and Rice T M 1987 Phys. Rev. Lett. 59 2772

[74] Thomsen C, Cardona M, Kress W, Liu R, Genzel L, Bauer M, Schönherr E and Schröder U 1988 Solid State Commun. 65 1139

[75] Ruf T, Thomsen C, Liu R and Cardona M 1988 Phys. Rev. B 38 11985

[76] Klein M V and Dierker S B 1984 Phys. Rev. B 29 4976

[77] Zeyher R and Zwicknagl G 1988 Solid State Commun. 66 617

[78] Zeyher R and Zwicknagl G 1990 Z. Phys. B 78 175

[79] Nicol E J, Jiang C and Carbotte J P 1993 Phys. Rev. B 47 8131

[80] Devereaux T P 1994 Phys. Rev. B 50 10287

[81] Macfarlane R M, Krantz M C, Rosen H J and Lee V Y 1989 Physica C 162 1091

[82] Krantz M C, Rosen H J, Macfarlane R M and Lee V Y 1988 Phys. Rev. B 38 4992

[83] Hadjiev V G, Thomsen C, Erb A, Muller-Vogt G, Koblischka M R and Cardona M 1991 Solid State Commun. 80 643

[84] Thomsen C, Friedl B, CieplakMand CardonaM 1991 Solid State Commun. 78 727

[85] Greaves C and Slater P R 1989 Supercond. Sci. Technol. 2 5

[86] Fisher B, Genossar J, Kuper C G, Patlagan L, Reisner G M and Knizhnik A 1993 Phys. Rev. B 47 6054

[87] Gunasekaran R A, Ganguly R and Yakhmi J V 1995 Physica C 243 160

[88] Varlamov S and Seibold G 2002 Phys. Rev. B 65 132504

[89] Limonov M F, Tajima S and Yamanaka A 2000 Phys. Rev. B 62 11859

[90] Strohm T, Belitsky V I, Hadjiev V G and Cardona M 1998 Phys. Rev. Lett. 81 2180

[91] Bock A, Ostertun S, Das Sharma R, Rübhausen M, Subke K O and Rieck C T 1999 Phys. Rev. B 60 3532

[92] Misochko O V, Kuroda K and Koshizuka N 1997 Phys. Rev. B 56 9116

[93] Heyen E T, Cardona M, Karpinski J, Kaldis E and Rusiecki S 1991 Phys. Rev. B 43 12958

[94] McCarty K F, Radousky H B, Liu J Z and Shelton R N 1991 Phys. Rev. B 43 13751

[95] Burns G, Chandrashekhar G V, Dacol F H and Strobel P 1989 Phys. Rev. B 39 775

[96] Bokholt M, Erie A, Splittgerber-Hünnekes P C and Güntherodt G 1990 Solid State Commun. 74 1107

[97] Martin A A, Sanjurjo J A, Hewitt K C, Wang X Z, Irwin J C and Lee M J G 1997 Phys. Rev. B 56 8426

[98] Martin A A and Lee M J G 1995 Physica C 254 222

[99] Leach D H, Thomsen C and Cardona M 1993 Solid State Commun. 88 457

[100] Krantz M C, Thomsen C, Mattausch Hj and Cardona M 1994 Phys. Rev. B 50 1165

[101] Zhou X J, Cardona M, Colson D and Viallet V 1997 Phys. Rev. B 55 12770

[102] Wagner J L, Hunter B A, Hinks D G and Jorgensen J D 1995 Phys. Rev. B 51 15407

[103] Hadjiev V G, Zhou X J, Strohm T, Cardona M, Lin Q M and Chu C W 1998 Phys. Rev. B 58 1043

[104] Allen P B 1972 Phys. Rev. B 6 2577

[105] Hadjiev V G, Cardona M, Du Z L, Xue Y Y and Chu C W 1998 Phys. Stat. Sol. (b) 205 R1

[106] Li R, Weimer U, Feile R, Tomé-Rosa C and Adrian H 1991 Physica C 175 89

[107] Sherman E Y, Li R and Feile R 1995 Phys. Rev. B 52 R15757

[108] Sherman E Y and Misochko O V 2001 Phys. Rev. B 63 104520

[109] Friedl B, Thomsen C, Habermeier H U and Cardona M 1991 Solid State Commun. 78 291

[110] Ostertun S, Kiltz J, Bock A, Merkt U andWolf T 2001 Phys. Rev. B 64 064521

[111] Zhao S C, Hou D, Wu Y, Xia T L, Zhang A M, Chen G F, Luo J L, Wang N L, Wei J H, Lu Z Y and Zhang Q M 2009 Supercond. Sci. Technol. 22 015017

[112] Gallais Y, Sacuto A, Cazayous M, Cheng P, Fang L andWen H H 2008 Phys. Rev. B 78 132509

[113] Okazaki K, Sugai S, Niitaka S and Takagi H 2011 Phys. Rev. B 83 035103

[114] Um Y J, Park J T, Min B H, Song Y J, Kwon Y S, Keimer B and Le Tacon M 2012 Phys. Rev. B 85 012501

[115] Chaudi`ere L, Gallais Y, Cazayous M, M´easson M A, Sacuto A, Colson D and Forget A 2011 Phys. Rev. B 84 104508

[116] Boeri L, Calandra M, Mazin I I, Dolgov O V and Mauri F 2010 Phys. Rev. B 82 020506

[117] Litvinchuk A P, Hadjiev V G, Iliev M N, L¨u B, Guloy A M and Chu C W 2008 Phys. Rev. B 78 060503

[118] Choi K Y, Wulferding D, Lemmens P, Ni N, Bud'ko S L and Canfield P C 2008 Phys. Rev. B 78 212503

[119] Hu W Z, Dong J, Li G, Li Z, Zheng P, Chen G F, Luo J L and Wang N L 2008 Phys. Rev. Lett. 101 257005

[120] Chauvi'ere L, Gallais Y, Cazayous M, Sacuto A,M´eassonMA, Colson D and Forget A 2009 Phys. Rev. B 80 094504

[121] Rahlenbeck M, Sun G L, Sun D L, Lin C T, Keimer B and Ulrich C 2009 Phys. Rev. B 80 064509

[122] Boeri L, Dolgov O V and Golubov A A 2008 Phys. Rev. Lett. 101 026403

[123] Yildirim T 2009 Phys. Rev. Lett. 102 037003

[124] Yildirim T 2009 Physica C 469 425

[125] Litvinchuk A P, L¨u B and Chu C W 2011 Phys. Rev. B 84 092504

[126] Choi K Y, Lemmens P, Eremin I, Zwicknagl G, Berger H, Sun G L, Sun D L and Lin C T 2010 J. Phys.: Condens. Matter 22 115802

[127] Um Y J, Subedi A, Toulemonde P, Ganin A Y, Boeri L, Rahlenbeck M, Liu Y, Lin C T, Carlsson S J E, Sulpice A, Rosseinsky M J, Keimer B and Le Tacon M 2012 Phys. Rev. B 85 064519

[128] Gnezdilov V, Pashkevich Y, Lemmens P, Gusev A, Lamonova K, Shevtsova T, Vitebskiy I, Afanasiev O, Gnatchenko S, Tsurkan V, Deisenhofer J and Loidl A 2012 Phys. Rev. B 83 245127

[129] Khasanov R, Bendele M, Amato A, Babkevich P, Boothroyd A T, Cervellino A, Conder K, Gvasaliya S N, Keller H, Klauss H H, Luetkens H, Pomjakushin V, Pomjakushina E and Roessli B 2009 Phys. Rev. B 80 140511

[130] Stock C, Rodriguez E E, Green M A, Zavalij P and Rodriguez-Rivera J A 2011 Phys. Rev. B 84 045124

[131] Zhang A M, Liu K, Xiao J H, He J B, Wang D M, Chen G F, Normand B and Zhang Q M 2012 Phys. Rev. B 85 024518

[132] Zhang Y, Yang L X, Xu M, Ye Z R, Chen F, He C, Xu H C, Jiang J, Xie B P, Ying J J,Wang X F, Chen X H, Hu J P, Matsunami M, Kimura S and Feng D L 2011 Nat. Mater. 10 73

[133] Rettig L, Cort és R, Jeevan H S, Gegenwart P,Wolf T, Fink J and Bovensiepen U 2013 arxiv:1304.5355V1 [cond-mat.supr-con]

[134] Mansart B, Boschetto D, Savoia A, Rullier-Albenque F, Bouquet F, Papalazarou E, Forget A, Colson D, Rousse A and Marsi M 2010 Phys. Rev. B 82 024513

[135] Miao R D, Bai Z, Yang J, Chen X, Cai D, Fan C H, Wang L, Zhang Q L and Chen L A 2013 Solid State Commun. 154 11

[136] Luo C W, Wu I H, Cheng P C, Lin J Y, Wu K H, Uen T M, Juang J Y, Kobayashi T, Chareev D A, Volkova O S and Vasiliev A N 2013 J. Supercond. Nov. Magn. 26 1213

[137] Luo C W, Wu I H, Cheng P C, Lin J Y, Wu K H, Uen T M, Juang J Y, Kobayashi T,Wen Y C, Huang TW, Yeh KW,Wu M K, Chareev D A, Volkova O S and Vasiliev A N 2012 New J. Phys. 14 103053

[138] Subedi A, Zhang L J, Singh D J and Du M H 2008 Phys. Rev. B 78 134514

[139] Bazhirov T and Cohen M L 2012 Phys. Rev. B 86 134517

[140] Li J and Huang G Q 2013 Solid State Commun. 159 45

[141] Bhoi D, Mandal P and Choudhury P 2008 Supercond. Sci. Technol. 21 125021

[142] Mertelj T, Kusar P, Kabanov V V, Stojchevska L, Zhigadlo N D, Katrych S, Bukowski Z, Weyeneth S, Karpinski J and

Mihailovic D 2010 Phys. Rev. B 81 224504

[143] Kordyuk A A, Zabolotnyy V B, Evtushinsky D V, Kim T K, Morozov I V, Kulic M L, Follath R, Behr G, Buchner B and Borisenko S V 2011 Phys. Rev. B 83 134513

[144] Jishi R A and Alyahyaei H M 2010 Adv. Condens. Matter Phys. 2010 804343

[145] Cohen R E, Pickett W E and Krakauer H 1990 Phys. Rev. Lett. 64 2575

[146] Rodriguez C O, Liechtenstein A J, Mazin J J, Jepsen O, Andersen O K and Methfessel M 1990 Phys. Rev. B 42 2692

[147] Andersen O K, Liechtenstein A J, Rodriguez O, Mazin J J, Jepsen O, Antropov V P, Gunnarsson O and Gopala S 1991 Physica C 147 185

[148] Bohnen K P, Heid R and Krauss M 2003 Europhys. Lett. 64 104

[149] Dahm T, Manske D, Fay D and Tewordt L 1996 Phys. Rev. B 54 12006

[150] Pao C H and Schüttler H B 1998 Phys. Rev. B 57 5051

[151] Lanzara A, Bogdanov P V, Zhou X J, Kellar S A, Feng D L, Lu E D, Yoshida T, Eisaki H, Fujimori A, Kishio K, Shimoyama J I, Noda T, Uchida S, Hussain Z and Shen Z X 2001 Nature 412 510

[152] Lee J, Fujita K, McElroy K, Slezak J A, Wang M, Aiura Y, Bando H, Ishikado M, Masui T, Zhu J X, Balatsky A V, Eisaki H, Uchida S and Davis J C 2006 Nature 442 546

[153] Greco A and Zeyher R 1999 Phys. Rev. B 60 1296

[154] Bulut N and Scalapino D J 1996 Phys. Rev. B 54 14971

[155] Egami T 2005 Struct. Bond. 114 267

[156] Nazarenko A and Dagotto E 1996 Phys. Rev. B 53 R2987

[157] Kulić M L 2000 Phys. Rep. 338 1

[158] Blumberg G, Mialitsin A, Dennis B S, Zhigadlo N D and Karpinski J 2007 Physica C 456 75

[159] Liu A Y, Mazin I I and Kortus J 2001 Phys. Rev. Lett. 87 087005

[160] Wolf E L, Zasadzinski J, Arnold G B, Moore D F, Rowell J M and Beasley M R 1981 Phys. Rev. B 22 1214

[161] Klein B M, Boyer L L, Papaconstantopoulos D A and Mattheiss L F 1978 Phys. Rev. B 18 6411